\documentclass[%
mathleft,%
]{an}
\usepackage{graphicx}
\usepackage{times}
\begin{document}

\Pagespan{1}{}
\Yearpublication{2011}%
\Yearsubmission{2015}%
\Month{1}%
\Volume{999}%
\Issue{92}%

\title{Search for Extreme Rotation Measures in CSS Sources}

\author{W.\ D.\ Cotton\inst{1}\fnmsep\thanks{Corresponding author:
  \email{bcotton@nrao.edu}}\fnmsep
\and  E.\ Kravchenko\inst{2}
\and  Y.\ Kovalev\inst{2}
\and  E.\ Fomalont\inst{1}
}
\titlerunning{Extreme Rotation Measures}
\authorrunning{W.\ D.\ Cotton,\ E.\ Kravchenko,\  Y.\ Kovalov\ \& E.\ Fomalont}
\institute{National Radio Astronomy Observatory, 520 Edgemont Rd, Charlottesville, VA 22903, USA.
\and 
Lebedev Physical Institute, Profsoyuznaya 84/32, Moscow 117997, Russia}

\received{XXXX}
\accepted{XXXX}
\publonline{XXXX}

\keywords{CSS sources, Faraday rotation}

\abstract{%
Magnetized plasmas traversed by linearly polarized light will reveal
their presence by the frequency dependent Faraday rotation of the
angle of polarization.
The regions surrounding the black holes powering the jets in AGNs are
expected to have dense magnetized plasmas, possibly giving rise to
very large Faraday rotations.
Compact steep spectrum (CSS) sources are good candidates to search for
very large Faraday rotated components as they contain compact emission
from close to the black hole and many are strongly depolarized at
centimeter wavelengths as expected from strong Faraday effects.
We present data on several CSS sources (3C48, 3C138 and 3C147)
observed with the VLA at frequencies between 20 and 48 GHz in the most
extended configuration.
Large, but not excessive rotation measures are reported.
}
\maketitle

\section{Introduction}
High thermal electron densities and strong magnetic
fields are expected in the immediate vicinity of the black holes
powering AGN. 
These should introduce a large Faraday rotation of any polarized
emission passing through the region.  
On the other hand, many CSS sources show relatively low fractional
polarization and modest rotation measures. 
Cotton et al. (2003B) show that small CSS sources are
strongly depolarized at 1.4 GHz; presumably by a dense and variable
Faraday screen.  
Furthermore, Cotton et al. (2003A)  show that a component
in the CSS source 3C138 is seen moving behind a dense Faraday screen
with fine scale structure and an RM $\sim$5000 rad/m$^2$ through holes in this
screen.  
These argue that the low fractional polarization of CSS nuclei are due
to a dense Faraday depolarizing screen which might be revealed at
higher frequency.  
   We describe the results of examining high frequency (20-45 GHz)
EVLA data on a number of CSS sources looking for evidence of very large
rotation measures. 
The observations of 3C138 are particularly instructive as its core has
relatively low fractional polarization to at least 45 GHz and a
rotation measure comparable to that apparently seen through holes in
the Faraday screen at 5 GHz.  
Even higher frequencies may be needed to reveal a dense Faraday
screen.

\section {Faraday Rotation}
A magnetized plasma is birefringent and causes a rotation of the angle
of the plane of linear polarization ($\chi$) by the relationship:
$$
\chi\ = \chi_0 \ +\ RM\ \lambda^2
$$
where $\chi_0$ is the intrinsic angle, $RM$ is the ``Rotation
measure'' in radians m$^{-2}$ and $\lambda$ is the wavelength in m.
$RM$ is related to electron density ($n_e$) and component of the
magnetic field along the line of sight (${\bf B}_{||}$) by the
integral along the line of sight.
$$
RM\ =\ const \int n_e {\bf B}_{||} dl
$$

\section {Faraday Rotation in AGNs}
\begin{figure*}
\includegraphics[width=6.5in]{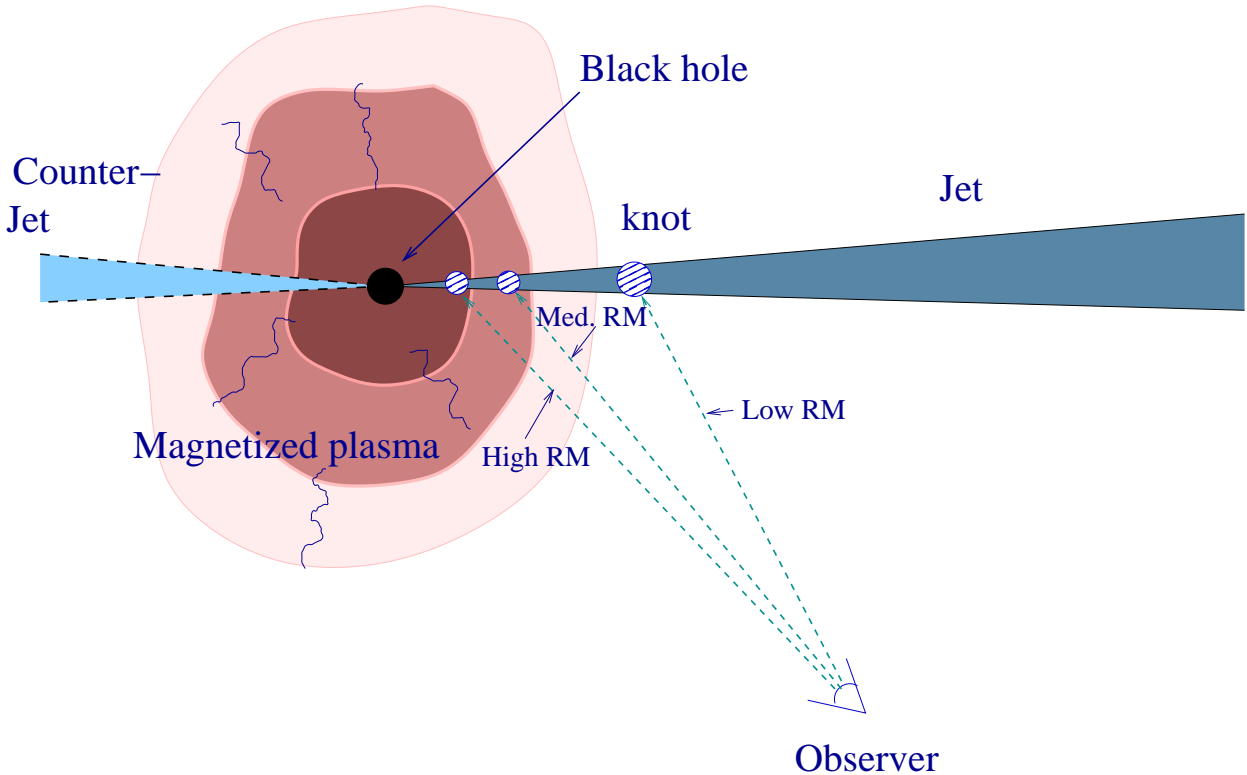}
\caption{ 
Cartoon of the inner region of a radio jet showing a potential case of
multiple components with different rotation measures.
} 
\label{cartoon}
\end{figure*}

AGNs are expected to have dense magnetized plasmas in the region of the
black hole and synchrotron emission from the jet is intrinsically
linearly polarized.
Figure \ref{cartoon} illustrates a potential case where multiple knots
in a radio jet have different Faraday rotations.
In the case of widely varying Faraday rotation in front of various
portions of the source, the polarization have wash out, strongly
depolarizing the overall emission.
Strong depolarization is frequently seen near AGN nuclei, especially
at long wavelengths, which may be the result of such Faraday
depolarization. 

Since Faraday effects depend strongly on the wavelength, observations
at shorter wavelengths will be less effected by Faraday effects.
Components which are Faraday depolarized at longer wavelengths may
become less depolarized resulting in higher fractional polarization
and potentially higher Faraday Rotation.

Another argument for observations at shorter wavelength is that the
synchrotron opacity increases towards the black hole in the inner part
of the jet.
Shorter wavelengths probe regions closer to the nucleus with
potentially higher rotation measures.

The prototype for very high rotation measures near an AGN is Sgr A*
for which values of $\sim 500,000$ rad m$^{-2}$ have been seen
(Bower et al. 2003).
In this source there is no strong jet with lower rotation measure to
complicate the observation.
However, as AGNs go, this is fairly weak; bigger AGNs may have
emission with much higher rotation measures.

\section {Data and processing}
The CSS sources 3C48, 3C138, 3C147 and 3C286 are included in the EVLA
standard calibrator list and were observed in the most extended,
``A'', configuration Feb. 16-17, 2014.
Each source was observed cyclically in each receiver band in a 36 hour
session giving excellent uv coverage.
The range 20 - 45 GHz (highest frequencies available in the VLA) was
covered with the K, Ka, and Q band receivers. 
Frequency coverage consisted of 2 $\times$ 1 GHz subbands in each
receiver band.
Since 3C286 was the primary calibrator, it was not included as a
potential target.
Calibration and imaging used the Obit package (Cotton 2008). \footnote{http://www.cv.nrao.edu/$\sim$bcotton/Obit.html}
Calibration followed the usual practices with 3C286 as the flux
density, cross polarized delay and polarization angle calibrator.
The combination of the flat spectrum quasars 0217+7349, J1153+8058,
and J1800+7828 were used to determine the instrumental calibration.
These circumpolar sources were each well sampled over the full
360$^\circ$ range of parallactic angle during the 36 hour observing
session allowing good separation of source and instrumental
polarization.

As the target sources were all resolved, a combination of Briggs
robustness parameter and uv taper were used to produce images at a
common resolution in each receiver band.
Each 1 GHz subband was broken into 8 frequency bins and imaging used
the Obit wideband imager MFImage.
In addition, full resolution images were obtained at 43 GHz (Q band).

\section {Results}
Full resolution images of the CSS sources are shown in Figures
\ref{3C48QBand}, \ref{3C138QBand} and \ref{3C147QBand}.
These give Stokes I as contours and linear polarization as vectors and
grayscale. 
The EVPA and fractional polarizations as a function of wavelength
squared are given in Figures \ref{3C48RM}, \ref{3C138RM} and
\ref{3C147RM}. 
The fitted core rotation measured and errors are shown in Table \ref{RMs}.
\begin{table}
\caption{Core Rotation Measures}
\label{RMs}
\begin{tabular}{|r|r|r|}   \hline
\hline
 Source &  RM & $\pm$   \\
      &     rad\ m$^{-2}$ & rad\ m$^{-2}$  \\
\hline
3C48  & 9671    & 484 \\
3C138 & -3224   & 101  \\
3C147 &  -1509  &  7 \\
\hline
\end{tabular}
\end{table}

\subsection{3C48}
\begin{figure}
\includegraphics[width=2.5in]{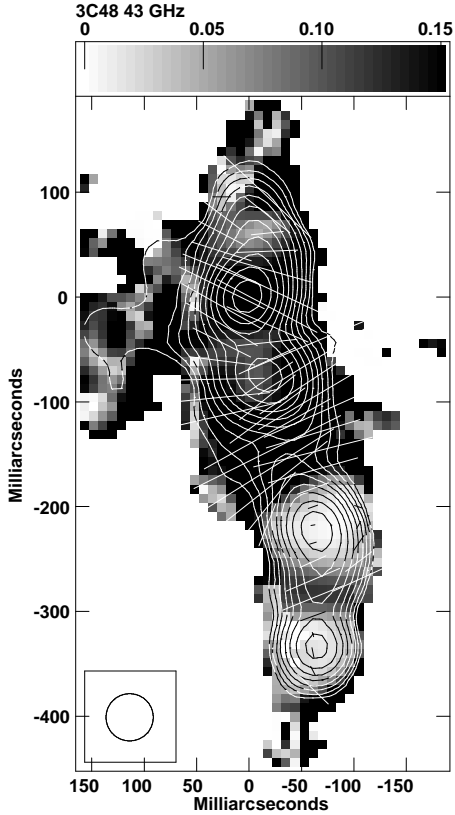}
\caption{ 
3C48 at 43 GHz, Stokes I is shown in contours, linear polarization E
vectors are given as lines and the fractional polarization is given as
the inverted gray scale with a scale bar at the top.
Contours are at powers of $\sqrt{2}$ times 2 mJy/beam.
The circle in the lower left corner gives the size of the resolution
and the ``core'' is the bottom most.
} 
\label{3C48QBand}
\end{figure}

\begin{figure}
\includegraphics[width=2.6in,angle=-90]{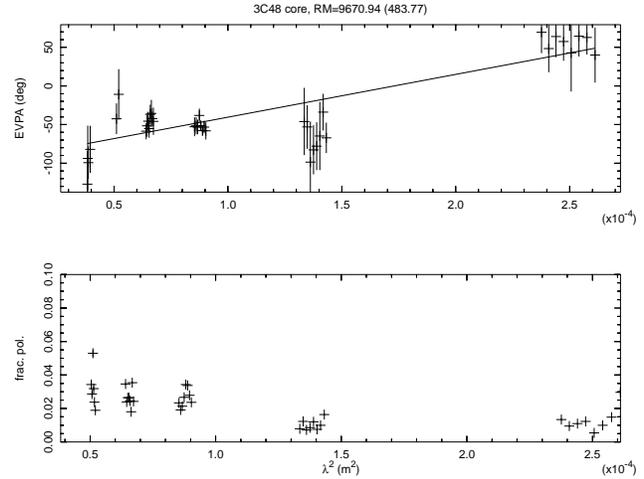}
\caption{ 
3C48 core:
top plot gives the EVPA as a function of wavelength squared and the
line the fitted rotation measure.
Bottom plot gives the fractional linear polarization as a function of
wavelength squared. 
} 
\label{3C48RM}
\end{figure}

Data for 3C48 are shown in Figures \ref{3C48QBand} and \ref{3C48RM}.
The fractional polarization of the core is fairly low but increases
towards shorter wavelengths.
The fitted RM ($\sim 10,000$) is relatively high but the significant
scatter about the fit suggests more complex Faraday effects 
(Sokoloff et al. 1998).

\subsection{3C138}
\begin{figure}
\includegraphics[width=3.2in]{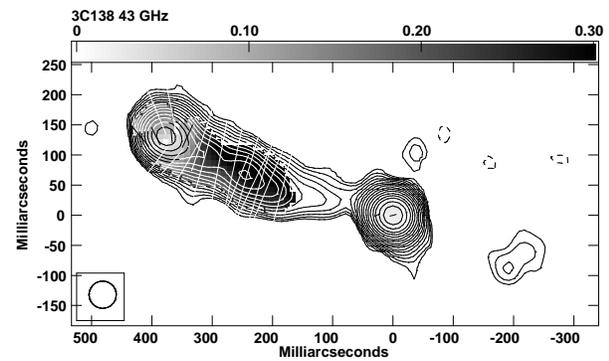}
\caption{ 
Like Figure \ref{3C48QBand} but for 3C138.
Contours are at powers of $\sqrt{2}$ times 1 mJy/beam
and the ``core'' is the strong component at position (0,0).
} 
\label{3C138QBand}
\end{figure}

\begin{figure}
\includegraphics[width=2.6in,angle=-90]{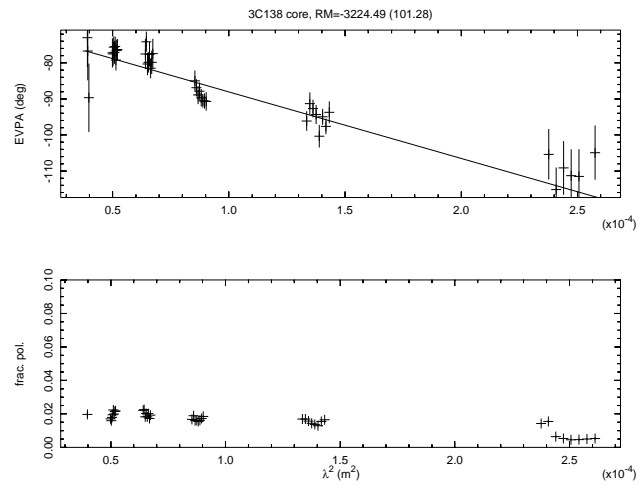}
\caption{ 
3C138 core: like Figure \ref{3C48RM} but for 3C138.
} 
\label{3C138RM}
\end{figure}

Data for 3C138 are shown in Figures \ref{3C138QBand} and
\ref{3C138RM}.
3C138 has a low fractional polarization to the shortest wavelengths
shown in Figure \ref{3C138RM}.
The fitted RM ($\sim -3,000$) is less than the ($\sim -5,000$) reported
in  Cotton et al. (2003A)  at 6 cm wavelength with data
that were interpreted as seeing a moving component behind a dense
Faraday screen.

\subsection{3C147}
\begin{figure}
\includegraphics[width=2.5in]{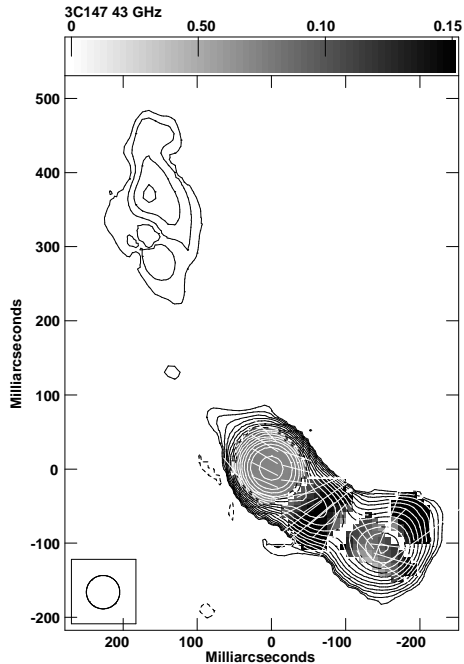}
\caption{ 
Like Figure \ref{3C48QBand} but for 3C147.
Contours are at powers of $\sqrt{2}$ times 1 mJy/beam
and the ``core'' is the strong component at position (0,0).
} 
\label{3C147QBand}
\end{figure}

\begin{figure}
\includegraphics[width=2.6in,angle=-90]{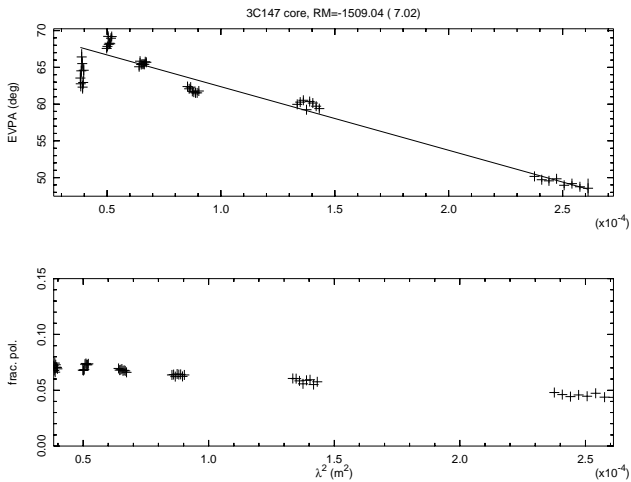}
\caption{ 
3C147 core: like Figure \ref{3C48RM} but for 3C147.
} 
\label{3C147RM}
\end{figure}

Data for 3C147 are shown in Figures \ref{3C147QBand} and
\ref{3C147RM}.
At cm wavelengths this source is strongly depolarized but in Figure
\ref{3C147RM} the core has relatively high fractional polarization but
a low RM ($\sim -1,500$)

\section{Discussion}
None of the sources examined here show the very large ($>$100,000 rad
m$^{-2}$) rotation measures expected very close to the black holes.
There are a number of potential causes for this.
First, the observed wavelengths may be long enough that the source is
opaque sufficiently close to the center to inhibit illuminating the
densest plasma.
Another possibility is that the emission is dominated by components
far enough down the jet to have only relatively low RM foregrounds.
In the latter case, there may be multiple components with different RMs.

\section{Further Work}
Observations at even shorter wavelengths provide a combination of
probing closer to the nucleus and reduced sensitivity to Faraday
effects and may allow detection of very large RMs.

In order to test for the presence of weak, high RM components in the
presence of strong, low RM components, the Faraday analysis technique
(Brentjens and de Bruyn, 2005) may be helpful.
This technique is to Fourier transform the polarization data in
$\lambda ^2$ space and look for multiple components; the Fourier
transform will separate the different periodic behaviors in 
$\lambda ^2$.
For this technique to work well, relatively dense coverage in $\lambda
^2$ is desirable.
We are currently engaged in a VLA project covering most of the
frequencies from 20 to 45 GHz on a number of bright AGN.

\acknowledgements
The National Radio Astronomy Observatory (NRAO) is operated by
Associated Universities Inc., under cooperative agreement with the
National Science Foundation.
EVK was suppported in part by the Russian Foundation for basic
Research (project 14-02031789 mol\_a).
%
%

\end{document}